\documentclass[runningheads]{llncs}
\usepackage{subfigure}
\usepackage[T1]{fontenc}
\usepackage{amsfonts}
\usepackage{mathrsfs}
\usepackage{amsmath}
\usepackage{graphicx}
\usepackage{multirow}
\begin{document}
\title{DSCom: A Data-Driven Self-Adaptive Community-Based Framework for Influence Maximization in Social Networks\thanks{Supported by organization x.}}
%
%
\author{Yuxin Zuo \and
Haojia Sun \and
Yongyi Hu \and
Jianxiong Guo \and
Xiaofeng Gao}
\authorrunning{F. Author et al.}
%
\institute{Shanghai JiaoTong University, CN}
\maketitle              
\begin{abstract}

Influence maximization aims to find a subset of seeds that maximize the influence spread under a given budget. In this paper, we mainly address the data-driven version of this problem, where the diffusion model is not given but needs to be inferred from the history cascades. Several previous works have addressed this topic in a statistical way and provided efficient algorithms with theoretical guarantee. However, in their settings, though the diffusion parameters are inferred, they still need users to preset the diffusion model, which can be an intractable problem in real-world practices. In this paper, we reformulate the problem on the attributed network and leverage the node attributes to estimate the closeness between the connected nodes. Specifically, we propose a machine learning-based framework, named DSCom, to address this problem in an heuristic way. Under this framework, we first infer the users' relationship from the diffusion dataset through attention mechanism and then leverage spectral clustering to overcome the influence overlap problem in the lack of exact diffusion formula. Compared to the previous theoretical works, we carefully designed empirical experiments with parameterized diffusion models based on real-world social networks, which prove the efficiency and effectiveness of our algorithm.

\end{abstract}
\section{Introduction}
With the popularization of the social network, crucial information is spread in a more cost-effective way. Taking advantage of the gigantic social media data, governments promote political advocacy, companies advertise products, and people share their opinions. This leads to the study of the \emph{Influence Maximization} (IM) problem, aiming at selecting a few influentials in the social network to spread information based on the "word-of-mouth" strategy. There exist many applications regarding the IM problem, such as viral marketing~\cite{chen2010scalable}, and network monitoring~\cite{gomez2012inferring}.

In the seminal paper, Kempe et al.~\cite{KempeKT03} first modeled IM as a combinatorial optimization problem where the diffusion process should be manually predefined. Independent Cascade (IC) and Linear Threshold (LT) model are two diffusion patterns firstly proposed in~\cite{KempeKT03} and then widely adopted as the evaluation metric for various IM algorithms. However, the approach of pre-defining diffusion models is disadvantageous and impractical in application scenarios, because in reality, the true diffusion mechanism is scarcely given as a directly observable information. It may vary according to topics, location, time and other related variables. This problem has been noticed by the research community and usually referred to as network inference, where researchers infer the diffusion models from the observed cascade samples. Network inference was originally proposed to reconstruct the edge set such that the network structure can best explain the observed infection times~\cite{gomez2012inferring,myers2010convexity}. Recently, researches have reformulated this problem under the IM setting to infer the diffusion parameters under some predefined diffusion models and they leverage some statistical techniques to conceive inference algorithms with theoretical guarantees~\cite{chen2021network,onlineIM21}. However, all existing algorithms are restricted to specific diffusion models, which are rarely given as directly observable information in practice. In order to design a practical IM algorithm generalizable to various diffusion patterns, in this paper, we introduce a novel formulation of data-driven IM problem. 

\begin{figure}[!t]
	\centering
	\includegraphics[width=\columnwidth]{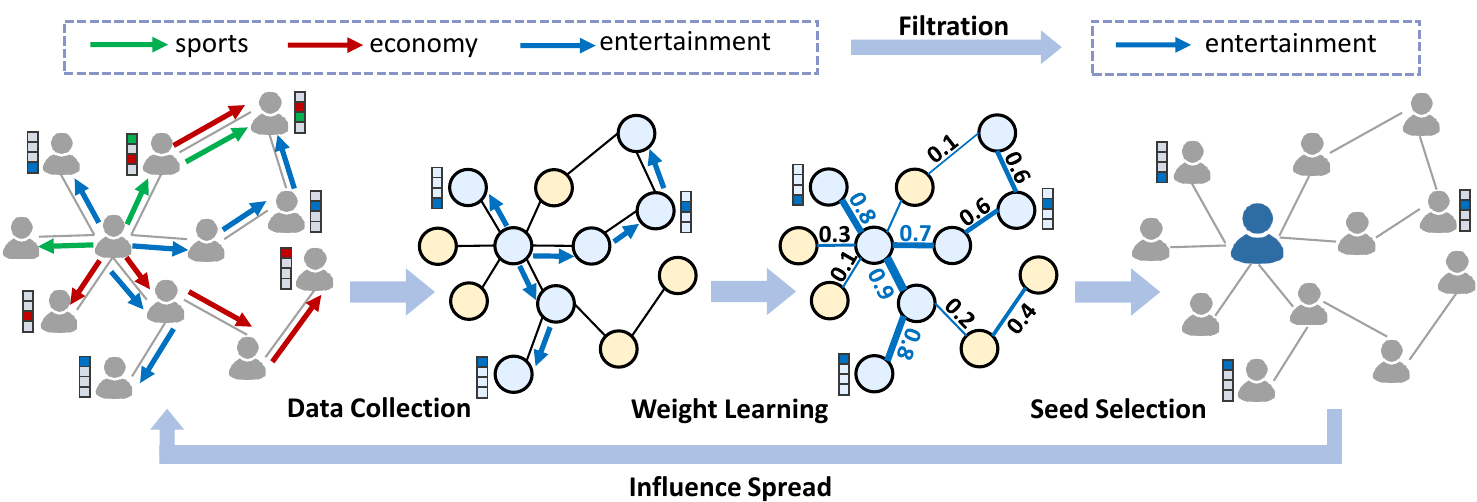}
	\caption{Data-driven IM Illustration. First, social network is modeled into an attributed graph and collect propagation trace with our desired filtration, according to topic for example. Then with the help of the collected dataset we evaluate the connection strength between each neighbors and augment the original graph into a weighted one. Finally, the seed set is selected to maximize the influence spread on the given network.}
	\label{fig: application}
\end{figure}

\textbf{Data-driven IM problem} takes advantage of the historic diffusion cases to improve the influence performance. In contrast to the difficulty of obtaining the diffusion model along with its parameters, the successful diffusion history is usually easy to collect, even on a large scale. For example, if a user $v$ retweet a piece of information published by user $u$, it implies that user $u$ influences user $v$ and the edge $(u,v)$ can be recorded as a successful diffusion case. Under this formulation, we make a basic assumption that there exists an underlying diffusion model, who determines the distribution of our diffusion dataset. \textbf{It should be clearly stated that the exact mathematical formula of the underlying diffusion model can not be directly obtained in any way, while the sampling cascades conforms to its distribution.} An illustration of our data-driven IM problem is presented in Fig.~\ref{fig: application}.

Selecting seed nodes based on the diffusion dataset is a non-trivial problem. To tackle this problem, we design a Machine Learning (ML)-based heuristic method: \textbf{Data-driven Self-Adaptive Community-based (DSCom)} framework, which makes effort to re-construct the relationship among users in social networks and perform seed selection in a community-based way. Sequentially, our DSCom framework can be divided into three modules: relation learning, community discovery, and seed selection. First, in relation learning, we transform the target network to node embedding through multi-head Graph Attention Network (GAT), which is trained by the diffusion dataset to minimize a loss function based on the posterior probability. We then extract the learned attention functions to evaluate the connection strength for each edge in the graph. Second, in community discovery, with the weighted graph augmented in the previous step, we adopt a normalized spectral clustering algorithm to achieve community partition. The basic intuition behind this community-based approach is to avoid the influence overlap problem by sparsely separate the seed nodes. Third, after obtaining the partitioned communities, we use a centrality-based method to select very few seed nodes in each community. Finally, concerning the experiments, while the previous works fail to provide empirical studies, we introduce the parameterized diffusion models and based on which we conduct a group of experiments to verify our proposed framework by comparing the influence performance with other baseline algorithms.

Our main contributions can be summarized as follows: 

\begin{itemize}
    \item \textbf{Problem Formulation.} We address a more practical formulation of data-driven IM problem, where the type of diffusion model is not given in advance.
    \item \textbf{Novel Pipeline.} A ML-based heuristic framework, DSCom, is proposed to tackle the novel problem, integrating graph attention network (GAT) and NCut technique with the novel idea of attention extraction.
    \item \textbf{Experimental Results.} Compared to the lack of empirical study of the previous works, we design and conduct the empirical experiments and it proves the effectiveness and efficiency of our pipeline.
\end{itemize}


\section{Related Work}
\label{sec:related}

\textbf{Network Inference in Influence Maximization.} The IM problem was first proposed by Kempe et al.~\cite{firstIM}. In this seminal paper, authors proved that the IM problem is NP-hard and gave a greedy algorithm with the theoretical guarantee of $(1-1/e)$ approximation. In the past two decades, researchers have proposed either heuristic algorithms, for example, LDAG~\cite{chen2010scalable} and SIMPath~\cite{goyal2011simpath} or approximate algorithms, for instance, IMM~\cite{IMM}, SSA/DSSA~\cite{ssa}, and OPIM-C~\cite{OPIM-C}, to tackle this problem. With the rising of deep learning techniques, a new trend for tackling IM problem based on machine learning has emerged, e.x., MAIM~\cite{MAIM}. However, in the canonical setting, it is assumed that the diffusion model is given, which is rarely the case in real world practices. Therefore, researchers combine network inference with the IM problem, where the diffusion parameters are not directly given but supposed to be inferred from the diffusion cascades~\cite{chen2021network,onlineIM21}. Statistical techniques are typically applied to bound the estimation error and deduce the approximate rate.


\textbf{Graph Embedding.} It is difficult to solve complex large-scale network problems through traditional graph algorithms. Owing to the emergence of graph embedding, mining information in networks can be directly conducted in a low-dimensional vector space. The famous algorithm DeepWalk~\cite{deepwalk} adopted random walk to collect the information near nodes and exploited the skip-gram method to learn the nodes' representations. Besides, node2vec~\cite{node2vec}, NetMF~\cite{NetMF}, etc. were committed to improve the embedding quality. Recently, many studies utilized the graph neural network (GNN) to learn graph representations. Among them, the architectures that have received the most attention are the message-passing GNNs, such as Graph Convolutional Network (GCN)~\cite{GCN}, GraphSAGE~\cite{GRAPHSAGE}, and Graph Attention Network (GAT)~\cite{GAT}. The GNN aggregates the information of each node's neighbors at each layer, and in this way to learn effective representation by integrating topological information and local node features.


With the help of graph learning, we address a more practical formulation of data-driven IM problem on the attributed social network. Compared with the traditional network inference problem based on statistical technology, we do not depend on the given diffusion model, and we abandon the complicated mathematical deduction. They need a large number of data to bound the error, but we can achieve our goal, i.e., select the most influential seed set with only a small number of data, by efficiently mining network information.

\section{Problem Formulation}
\label{sec:problem}

In this section, we formally define our data-driven IM problem. To motivate this initiative, we first review the definition of the traditional IM problem. Then we point out two drawbacks brought by it in real applications. Finally, we reformulate it into a novel data-driven version to overcome these difficulties.

\subsection{Traditional IM Problem}
The IM problem is to select $k$ users who can achieve the maximum \emph{influence spread} across the network. To quantify this problem, we first formally introduce the \emph{Influence Spread} of a seed set. Given a graph $G=(V, E)$, a diffusion model $M$, and a seed set $S$, the influence spread (influence function) is the expected number of influenced users. It is denoted by $\sigma_{G,M}(\cdot)$ and defined as follows: $\sigma_{G,M}(S) = E[|I_M(S)|]$, where $\sigma_{G,M}: 2^{V} \rightarrow$ $\mathbb {R}_{\geq{0}}$ and $I_M(S)$ denotes the set of influenced users given seed set $S$ under diffusion model $M$. Then, the traditional IM problem can be defined as follows.
Given a positive integer $k$, a graph $G=(V, E)$, and a diffusion model $M$, the IM problem aims at selecting a node set $S^* \subseteq V$ such that $S^* = \mathrm{argmax} \left\{\sigma_{G,M}(S) \mid S \subseteq V, \lvert S \rvert \leq k\right\}$.

\subsection{Data-Driven IM Problem}
Due to the problem setting of traditional IM, \textbf{two major drawbacks} can seriously undermine the performance of the state-of-the-art algorithms in real-world scenarios. (1) The majority of the researches on the IM problem are targeted at specific diffusion models, such as IC, LT, etc. However, discovering the diffusion model with the observed cascades is usually an intractable problem in real-world practices. (2) Different application scenarios, such as topics or locations, generally do not share the same diffusion pattern. Thus, we cannot arbitrarily determine the diffusion parameters such as the influence probability in IC model.

Inspired by other successful data-driven approaches, such as recommendation system, we may also apply a similar idea to render our IM problem more practical by exploiting the diffusion dataset. Viewing the gigantic size of the mainstream online social networks and the rapid development of the data storage and processing capacity, we may safely conclude that the historical diffusion cases of relevant topics can be easily and massively collected by service providers. This lies a solid foundation for the proposition of data-driven version of IM problem, implicating its strong practical meaning and research interest.

We hereby formally formulate our data-driven IM problems defined on an attributed social network $G = (V, E, X)$, where $X$ is a $|V|\times F$ matrix containing the each user's personal features. To reduce the need of data for the estimation of influence one user can impose on another, we additionally take the users' attributes into consideration and assume that the diffusion process $M'_X$ is correlated with the users' attributes $X$, for simplicity, we denote it as $M'$ in the following part. We then define the form of diffusion dataset upon this network.

\begin{definition}[Diffusion Dataset]
    Given an attributed graph $G= (V, E, X)$ and an underlying diffusion model $M'$, a diffusion dataset $\mathbb{D}_{M'}$ is defined as a multi-set of diffusion pairs conforming to the distribution implied by $M'$:
    \begin{equation}
        \mathbb{D}_{M'} = \large\{ (u_i, v_i) \mid (u_i, v_i) \in E \large\}_{i=1}^{N} \sim \mathscr{D}_N(M')
        \label{eq2}
    \end{equation}
    where each element $(u_i, v_i)$ implicates that the node $u_i$ influenced $v_i$ once in the history, $N$ is the size of the dataset.
    \label{def:diffusion dataset}
\end{definition}

We would like to highlight the fact that the exact mathematical formulation of the diffusion model is hidden from the user, and therefore we need to exploit the multi-set $\mathbb{D}_{M'}$ to evaluate the diffusion process. To distinguish from the observable one in the traditional IM setting, previously noted as $M$, we choose $M'$ as the notation for our hidden diffusion process. Finally, our formulation of diffusion dataset is presented as follows.


\begin{definition}[Data-Driven IM]
    Given an attributed graph $G= (V, E, X)$, a diffusion dataset $\mathbb{D}_{M'}$, and a positive integer $k$, the data-driven IM problem aims at selecting a node set $S^*\subseteq V$ such that
    \begin{equation}
        S^* = \mathrm{argmax} \{\sigma_{G,{M'}}(S) \mid S \subseteq V, \lvert S \rvert \leq k\},
    \end{equation}
    where $\sigma_{G,{M'}}(\cdot)$ is the influence function under graph $G$ and underlying diffusion model $M'$, $\mathbb{D}_{M'}$ is the diffusion dataset generated from $M'$ defined in Def.~\ref{def:diffusion dataset}.
    \label{def:data IM}
\end{definition}


\section{DSCom Framework}
\label{sec:DSCOM}
\begin{figure*}[!t]
\centering
\includegraphics[width=\textwidth]{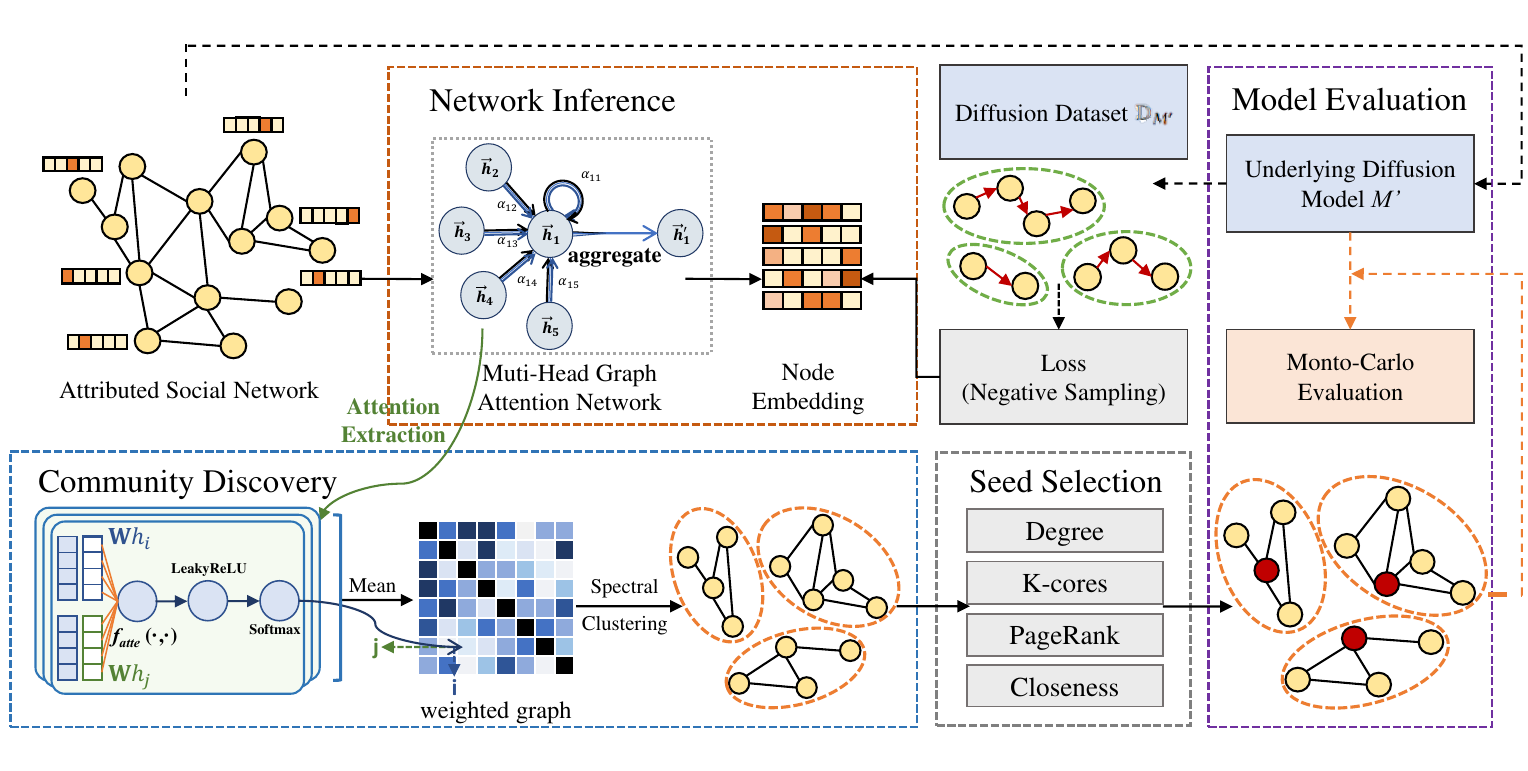}
\caption{DSCom overview. DSCom framework is composed of three sequential modules. The first module is named as relation learning, where we take advantage of users' features and the diffusion dataset to mine out the closeness of the relationship between neighbors. Specifically, the trained attention coefficients $\alpha_{ij}$ in the graph neural network are extracted to weight each edge in the graph. 
In the next module, spectral clustering algorithm is applied on the weighted graph for community discovery task. After the original network is divided into small communities, in the last module, we apply centrality-based measures to select the only few seed nodes from each sub-graph. Finally, the influence spread of the seed set can be evaluated through the Monte-Carlo simulation given the underlying diffusion model $M'$.}
\label{Fig:static}
\end{figure*}

To tackle the data-driven IM problem defined in Section ~\ref{sec:problem}, we hereby introduce our \textbf{Data-driven Self-Adaptive Community-based} (DSCom) framework, which is designed to be self-adaptive to different underlying diffusion models according to the observable diffusion chains. Our framework is mainly composed of three modules: Relation learning, Community Discovery, and Seed Selection. The overview of our framework is presented in Fig.~\ref{Fig:static}.

\subsection{Relation Learning}

The goal of this module is to discover the closeness of the relationship between each pair of neighbors. To achieve this objective, we apply the deep graph neural network, GAT, to integrate the structural information and node attributes, and then extract the attention function as the estimator of the neighbors' relationships among each other.

\noindent
\textbf{Graph Attention Network. }
It is complex to consider both the features of users and its local structure. Therefore, we use a deep neural network architecture to integrate the node attributes and the structural information. Graph Attention Network (GAT)~\cite{GAT} is a recently proposed technique that introduces the attention mechanism into Graph Convolutional Networks. As opposed to other message-passing GNNs, GAT is capable of assigning different levels of importance to different neighbors of a node, which is more in line with the real situation where people attach different importance to their neighbors.

For the consideration of being self-contained, we will briefly review the GAT model. The key idea behind GAT is to update the embedding of each node with that of its neighbors from the last layer in the form of weighted sum, where the weight is given by the attention function. Denote the embedding of node $u$ at layer $l$ as $h^{(l)}_i$, the attention function is defined as $ e_{ij}^{(l)}=f_{atte}\left(\mathbf{W}\mathrm{h}_i^{(l)}, \mathbf{W}\mathrm{h}_j^{(l)}\right) = \mathtt{LeakyReLU}\left({\mathbf{a}}^{T}\left[\mathbf{W}\mathrm{h}_i^{(l)} \| \mathbf{W}\mathrm{h}_j^{(l)}\right]\right)$,
where $f_{atte}(\cdot$, $\cdot)$ is the attention function, which can be implemented by a neural network with the weight vector $\mathbf{a} \in  \mathbb{R}^{2F^{\prime}}$ and a LeakyReLU function, while $ \cdot$ $\|$ $\cdot$ is the concatenation operator. Then, we normalize the coefficients of $e_{ij}^{(l)}$ using the softmax function to make them comparable across different nodes, and we denote the coefficient after softmax as $\alpha_{ij}^{(l)}$.

Finally, the embedding of node $u$ is updated as weighted sum, $h_{i}^{(l+1)} = \sum_{j\in\mathcal{N}_i}\alpha_{ij}^{(l)}h_j^{(l)}$, where $\mathcal{N}_i$ denotes the neighborhood of node $i$. In practice, to stabilize the learning process, Ashish Vaswani et al.~\cite{Attention} have found that it is beneficial to extend the mechanism by employing multi-head attention. 

\noindent
\textbf{Training Process. }
Aiming to learn the valid embedding of users, as well as the closeness between their neighbors, we exploit the diffusion dataset $\mathbb{D}_{M'}$ to construct diffusion chains from the successful diffusion pairs. For example, the pairs $(u, v)$ and $(v, w)$ can be concatenated into a diffusion chain $[u, v, w]$. 

We then train the network with the skip-gram architecture~\cite{skipgram}. Specifically, the probability that the user $v$ lies in the diffusion window of user $u$ is given by the softmax function.
\begin{equation}
    p(v \mid u; \theta) = {\exp(z_u \cdot z_v)}/{\sum\nolimits_{w \in V} \exp(z_u \cdot z_w)},\quad\forall u,v \in V
\end{equation}
where $z_u$ represents the node embedding of user $u$, and $\theta$ denotes the parameters of graph neural network.

Then we construct the loss function with probability multiplication for each $(u, v)$ pair co-occurring in the same diffusion chain in $\mathbb{D_{M'}}$ under the independence assumption. By taking log of the probability and applying negative sampling~\cite{negativesampling} to reduce the computational overhead, objective function is as follows:
\begin{equation}
    \underset{\theta}{\arg\max} \sum\nolimits_{(u,v)} \left[ \log \sigma(z_u \cdot z_v) + \sum\nolimits_{i=1}^K \log \sigma(-z_{u} \cdot z_{w_i})  \right]
\end{equation}
where $K$ is a positive integer larger than 1 indicating the ratio between negative and positive samples. Here we take $K\in[3,5]$. $(w_i)_{i=1}^K$ are nodes independently and uniformly sampled from the node set, and $\sigma(\cdot)$ denotes the sigmoid function.


\subsection{Community Discovery}

Without knowing the specific diffusion model, we heuristically propose dividing the network into communities before selecting the seed nodes so as to guarantee the sparsity of our seed set. In this section, we will first explain the intuition behind this idea, and afterwards, present our community discovery approach based on attention extraction and spectral clustering.

\begin{figure*}
\centering
\includegraphics[width=\linewidth]{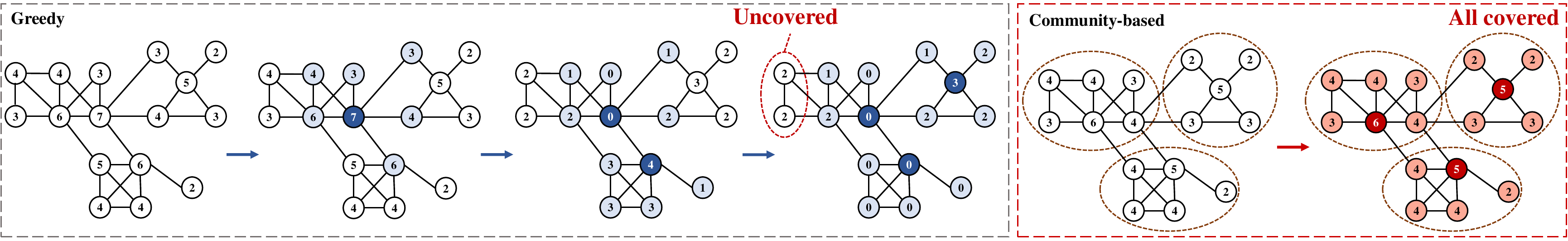}
\caption{Toy Example to motivate community detection. Given a graph $G$, we assume the selected seed node only influences its one-hop neighbors. The greedy algorithm aims at maximizing the marginal influence gain at each step (left). It fails to cover all nodes with a budget of three due to the influence overlap between selected seed nodes. While by pre-processing the graph through community detection before the selection process, this problem is successfully overcome (right).}
\label{Fig:toyexample}
\end{figure*}

\noindent
\textbf{Motivation:}
The traditional greedy algorithm adopts a point-by-point selection strategy, i.e., each selected node maximizes the marginal gain in the current state. However, two major issues remain to be solved. Firstly, as we have illustrated with a toy example in Fig.~\ref{Fig:toyexample}, the greedy algorithm could lead to the influence overlap between seed nodes. We will introduce the community detection as a heuristic approach to address this problem. Secondly and more importantly, under the data-driven IM, the underlying diffusion model $M'$ is unknown, which makes it intractable to estimate the influence spread of a given seed set.

Here, we adopt the community detection as a helper to our heuristic solution. This approach is well motivated for two reasons. (1) It is proved through empirical samples that the social network possesses significant and meaningful community structures~\cite{girvan2002community}. Therefore, the community detection result is expected to be of good quality. (2) Our approach is to select only few nodes within each community, so that the influence overlap can be minimized within the community. Meanwhile, due to the sparsity of the intra-community edges, the influence overlap problem can be further alleviated between communities.

\noindent
\textbf{Attention Extraction.}
Instead of performing the clustering task directly in the low-dimensional node embedding space, we choose to augment the network with the extracted attention coefficients $\alpha_{ij}$, then perform the community discovery task on the weighted graph. The advantage of this approach is two-fold: \textbf{(1) Relation Evaluation:} Same to the distance metrics in the latent space, attention function can also be used to evaluate the connection strength between the node pairs. \textbf{(2) Structure Preservation: } Comparing to node embeddings, our attention extraction approach preserves explicitly all the network edges. However, in the latent space, it is non-trivial to exactly reconstruct the edge set. 

\noindent
\textbf{Community Discovery.}
Many researchers have provided various algorithms to mine the underlying community structure behind the social network, such as hierarchical clustering~\cite{girvan2002community}, modularity maximization~\cite{modularity,modularityvariants}, statistical inference methods~\cite{karrer2011stochastic}, graph partitioning~\cite{Ncut}, etc. To select the best fit for our data-driven IM problem, we prefer they have two properties: (1) The number of the communities detected is flexible, meaning that given a positive integer $k$ as input, the community detection algorithm should return a result set with $k$ communities. The motivation is to be capable of adjusting the number of communities according to the given seed node budget, and there we can have very few seed nodes in each community to minimize the intra-community influence overlap. (2) The emergence of tiny communities should be avoided, because a very small community presents us with a dilemma: choosing some seed nodes in it, due to its relatively limited size, the influence could be trivial; not choosing any node, this community is then completely ignored from an algorithm design perspective, which may results in a sub-optimal solution. 

Although all the previously mentioned community detection algorithms have been proved empirically efficient in various real-world network structures, most of them either fail to give a community detection output for a given number of communities or they are unable to avoid the emergence of very small communities. To retain these two properties, we use the Normalized Cut~\cite{Ncut} as our optimization objective function for the community discovery module in our framework, and adopt the normalized spectral clustering algorithm.

\subsection{Seed Selection}

After we finish the community partition, seed selection is performed independently in each community with an assigned budget. Our strategy is to keep as few nodes as possible within each community such that the intra-community influence overlap would be minimized. As a consequence of not knowing the underlying diffusion model $M'$, it is impossible to evaluate the influence spread function $\sigma_{G,M'}$, or apply the greedy approach, i.e., to select the node that maximizes the marginal gain within the community becomes impractical. Hence, we adopt a heuristic approach to select the most appropriate node in the network structure. We compare several approaches to measure the centrality of a node through experiments, and these techniques can be listed as follows:
\begin{itemize}
    \item \textbf{Degree Centrality:} It assigns an importance score based on the number of edges connected by each node.
    \item \textbf{K-Core~\cite{kcore}:} K-core of a graph $G$ is the maximal sub-graph $H$ of $G$ such that the minimum degree of the nodes in $H$ is larger than $k$.
    \item \textbf{PageRank~\cite{pagerank}:} As a variant of Eigenvector Centrality, it assigns each nodes a score based on their connections and their connections of connections.
    \item \textbf{Closeness Centrality~\cite{closeness}:} It calculates the shortest paths among all nodes, then assigns each node a score based on its sum of the shortest paths.
\end{itemize}


For convenience, the DSCom with different seed selection strategies are sequentially named after their first letters as "D-D", "D-K", "D-PR" and "D-C".

\section{Experiments}
\label{sec:experiment}
In this section, we carefully designed various experiments under several real-world social network datasets to quantitatively evaluate our DSCom framework. Experiments are performed in a PC with an NVIDIA GeForce MX350 (8GB RAM), 16GB RAM and eight Intel(R) Core(TM) i7-1065G7 CPU @ 1.30GHz. The source code is available at https://github.com/annonymous-a/DSCom.

\subsection{Parameterized Diffusion Models and Datasets}

Traditionally, IC or LT with random parameters can be applied to empirically test the algorithms. However, these random diffusion models fail to take the user features into consideration to evaluate the edge weight. For instance, in IC model, two pairs of nodes share the same features may have completely different propagation probabilities if the edge weights are sampled in a completely random way. To tackle this problem, we introduce the parameterized diffusion model through integrating the node features into the diffusion pattern, it fits better the real-world diffusion mechanism. Here we first present the IC version in Def.~\ref{def:PIC}



\begin{definition}[Parameterized IC Model (PIC)]
    The diffusion process follows the IC model in~\cite{KempeKT03}, except that the influence probability $p_{u, v}$ is defined based on score function. Given two nodes $u \in V$, $v \in N_u$, and their node features $X_u$, $X_v$ respectively, the influence probability $p_{u, v}$ is defined $p_{u, v} = \sigma (a \cdot \mathrm{score}(X_u, X_v)+b)$, where $\mathrm{score}(X_u, X_v) = \mathbf{v}^T \mathrm{tanh}(\mathbf{W}[X_u \| X_v])$, $\mathbf{v}$ and $\mathbf{W}$ are random parameters used to generate different diffusion models. Also $\cdot \| \cdot$ indicates the concatenation operator and $\sigma(\cdot)$ is the sigmoid function.
    \label{def:PIC}
\end{definition}

The score function describes how much influence the user $u$ can impose on user $v$, and $a$ and $b$ are two linear transformation parameters. In addition, PIC Model only applies one layer of attention mechanism, but we can generalize it with more complex neural networks whenever necessary. Similarly, this formulation can be easily extended to LT diffusion model by defining the threshold between linked node pairs with Def.~\ref{def:PIC}.

\noindent
\textbf{Datasets.}
As defined in Sec.~\ref{sec:problem}, our algorithm takes two data sources as input: an attributed social network and a diffusion dataset. First, Our experiments are conducted on two different social networks --- Facebook \cite{Facebook} and Twitter \cite{Twitter}. These two datasets are both obtained from~\cite{data}, whose sizes are denoted by $(|V|,|E|)$. They are $(4039,88234)$ and $(10341, 505126)$. Due to the limitation of the computational resources, the experiments on much larger graphs are not feasible under current circumstances. We leave this scaling problem as a part of our future work and also welcome the research community to contribute.

In our data-driven IM problem, the underlying diffusion model $M'$ is unobservable. Therefore, the algorithm should exploit the successful diffusion cases to learn the propagation mechanism. In order to verify the performance of our algorithm, the experiments are designed as follows: we first pre-define a diffusion model $M'$ and then sample a diffusion dataset $\mathbb{D}_{M'}$ from it. We choose three different $M'$s: the traditional IC model, the PIC model and the LT model. In this way, after obtaining the seed set, we can evaluate its performance with the influence function induced by $M'$.

From a practical point of view, we attempt to generate a diffusion dataset that is a close approximation to what we may collect in the real-world scenario. The generation process is defined as follows: (1) Pick a small set of nodes $S$ from user set $V$ uniformly, simulating the situation that each person has the same probability to randomly come across the topic. (2) The influence diffuses through the network according to the underlying diffusion model $M'$. (3) Record the diffusion process into the diffusion dataset in the form of node pairs $(u,v)$, where node $v$ is activated by node $u$. In our experiment, we limit the size of the diffusion dataset to $1000$ to test the efficiency of our algorithm, even though a much larger dataset could be collected by mainstream online service providers.

\subsection{Performance Analysis}
In this section, we compare the performance of our DSCom framework with other IM algorithms in terms of the influence spread with different budgets and the running time. In our experiment, for DSCom, we choose the number of community equals the number of nodes selected, and each community get one seed node as budget. 

\noindent
\textbf{Baselines.}
Here we first present the baselines chosen for our experiments, basically two state-of-the-art algorithms and one novel deep reinforcement learning (DRL) based approach. They are (1) \textbf{Stop-and-Stare Algorithm (SSA)~\cite{NguyenDT18}} proposed by Nguyen et al., which is the state-of-the-art IM approximation algorithm, which provides a ratio of $(1-1/e-\epsilon)$. (2) \textbf{IMM~\cite{IMM}} improves over Reverse Reachable Sketch methods, TIM/TIM+~\cite{TIM} with a martingale-based approach. It provides a better analysis in the number of random RR sets required to ensure the same theoretical bound $1-1/e-\epsilon$. (3) \textbf{MAIM~\cite{MAIM}} is a recently proposed DRL algorithm to solve IM problem, aiming to explicitly solve the influence overlap problem.

\begin{table*}[!t]
\caption{Influence spread evaluation by $10$ times $1000$ MC for $20$ and $200$ seeds}
\centering 
\begin{tabular}{c|c|l|ccccc||ccccc}
\hline
\multirow{2}{*}{\#Seed\ } & \multicolumn{2}{c|}{Graph} & \multicolumn{5}{c||}{FB} & \multicolumn{5}{c}{TW} \\
\cline{2-13}
 & \multicolumn{2}{c|}{Methods} & IC & PIC1 & PIC2 & PIC3 & LT & IC & PIC1 & PIC2 & PIC3 & LT \\
\hline
\multirow{7}{*}{20} & \multirow{4}{*}{ $\mathbb{D}_{M'}$} & D-D & 129.2 & 658.7 & 953 & 3171 & 280.5 & \textbf{173.2} & 703.2 & 1274 & 4832 & 398.6 \\
& & D-K & 128.8 & 637.6 & 950.4 & \textbf{3173} & 248.8 & 168.6 & 712.4 & 1263 & 4757 & 374.7 \\
& & D-PR & \textbf{129.4} & 657.4 & \textbf{957.0} & 3172 & 277.5 & 170.6 & \textbf{720.9} & \textbf{1310} & \textbf{4903} & 399.4 \\
& & D-C & 122.8 & \textbf{659.4} & 952.8 & 3171 & \textbf{280.7} & 171.0 & 718.1 & 1302 & 4813 & \textbf{402.8} \\
\cline{2-13}
& \multirow{3}{*}{$M'$}
& SSA & 160.2 & 685.2 & 972.9 & 3151 & 326.3 & 205.7 & 728.3 & 1336 & 5205 & 451.7 \\
& & IMM & \textbf{165.1} & \textbf{689.7} & \textbf{975.0} & \textbf{3175} & \textbf{329.5} & \textbf{206.6} & \textbf{731.6} & \textbf{1339} & \textbf{5222} & \textbf{453.9} \\
& & MAIM & 132.8 & 607.3 & 873.7 & 2972 & 305.6 & 192.3 & 680.1 & 1249 & 4673 & 428.3 \\
\hline
\hline
\multirow{7}{*}{200} & \multirow{4}{*}{$\mathbb{D}_{M'}$} & D-D & 781.6 & 828.7 & 1168  & 3258 & 528.3 & 988.3 & 1058  & 2073 & 4910 & 1384  \\
& & D-K & \textbf{804.5} & 806.1 & 1153  & 3260 & 462.6 & 961.2 & 1023  & 2038 & 4987 & 1376  \\
& & D-PR & 802.4 & \textbf{831.7} & 1148  & 3259 & \textbf{536.4} & \textbf{987.6} & \textbf{1060} & \textbf{2114} & 5012 & \textbf{1417} \\
& & D-C & 786.2 & 822.9 & \textbf{1170} & \textbf{3262} & 528.4 & 970.4 & 1034 & 2086 & \textbf{5033} & 1393  \\
\cline{2-13}
& \multirow{3}{*}{$M'$} & SSA & 901.5 & 923.2 & 1217  & 3284 & 725.2 & 1186  & 1099  & 2201 & 5425 & 1883  \\
& & IMM & \textbf{908.4} & \textbf{926.6} & \textbf{1221} & \textbf{3286} & \textbf{732.2} & \textbf{1190} & \textbf{1103} & \textbf{2205} & \textbf{5432} & \textbf{1888} \\
& & MAIM & 811.3 & 817.6 & 1012  & 3118 & 673.0 & 1002  & 972.4 & 2043 & 4803 & 1721 \\
\hline
\end{tabular}
\label{table:InfluenceSpread}
\end{table*}

\noindent
\textbf{Influence Maximization. }
Influence spread is the most important indicator of the performance of our algorithms. Therefore, we conduct extensive experiments with different social networks, diffusion models, and budgets. Experimental results are presented in Table~\ref{table:InfluenceSpread}, which can be separated into two parts. One is labeled $M'$ representing the diffusion model is directly given as ground truth, the other is labeled $\mathbb{D}_{M'}$ meaning that the diffusion model should be inferred from diffusion dataset. The influence evaluation for the resulting seed set is executed by Monte-Carlo (MC) simulations. In this experiment, we use the average of $1000$ simulation times to evaluate the influence spread and repeat it $10$ times to test the standard deviation of our statistics. We observe that generally, the standard derivation is less than $2$ and in the most extreme case, less than $5$, showing that the chosen number of MC simulations is sufficient.

To clarify the abbreviations in Table~\ref{table:InfluenceSpread}, "PICx" indicates PIC model in Def.~\ref{def:PIC} with different random parameters. Experiments point out two essential comparison results. First, our DSCom is only around $10\%$ inferior against traditional approximation algorithms even though it lacks the key knowledge of underlying diffusion models. Second, based on our empirical evaluations, the D-PR can be the most robust version among its three counterparts, and therefore can be selected as the representative of our framework.

\begin{figure}
    \centering
    \subfigure[Seed node size $20$]{
        \includegraphics[width=0.41\textwidth]{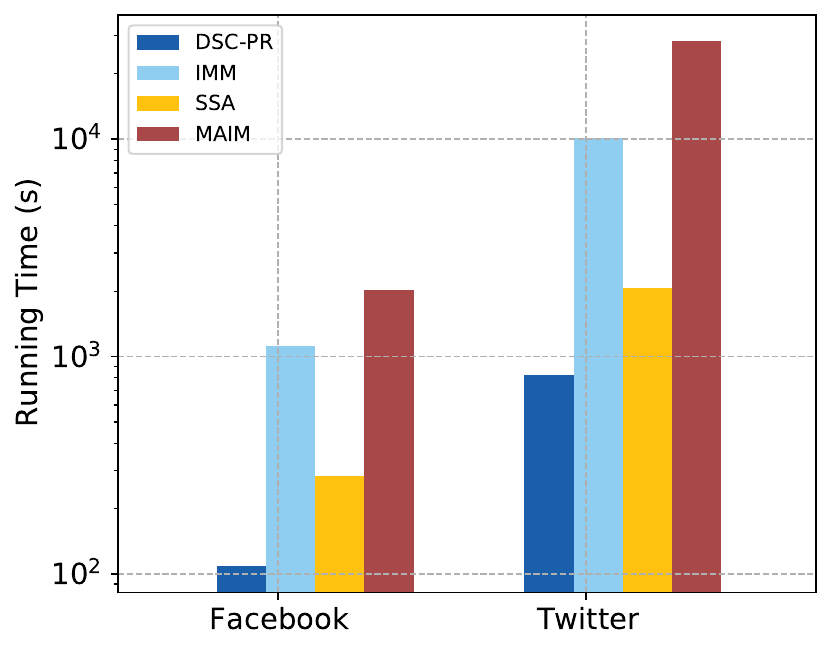}
        \label{fig:runningtime_1}
    }
    \subfigure[Seed node size $200$]{
        \centering
        \includegraphics[width=0.41\textwidth]{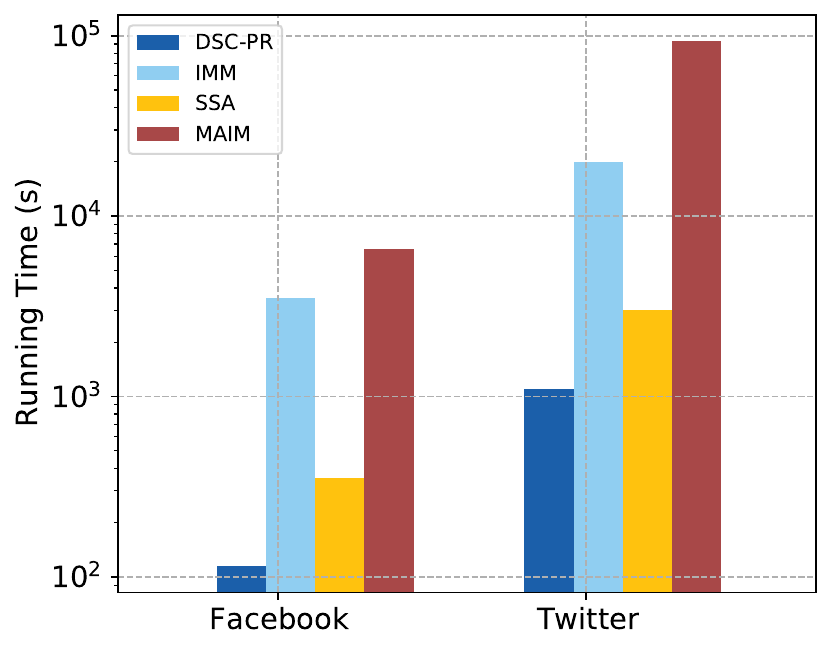}
        \label{fig:runningtime_2}
    }

    \caption{Computational time comparisons.}
    \label{fig: runningtime}
\end{figure}

\noindent
\textbf{Running Time.}
The running time for each algorithm presented in Table~\ref{table:InfluenceSpread} is noted according to these two datasets. To compare them in a more intuitive way, we take the average of the computational time over different settings on each graph and visualize them in Fig.~\ref{fig: runningtime}. It should be clarified that the training time of our relation learning module is not taken into account in the timing. It would be unfair to include the relation learning part into this comparison since the other models have already been given the correct diffusion model. For the record, the relation learning part of our algorithms for either of these two datasets costs less than $90$ minutes, which is acceptable and can definitely be improved with better computational resources.

Concerning the running time, our algorithm has an advantage over the other three methods in terms of the running time ($10\times$ faster than IMM, $3\times$ faster than SSA), especially when the seed set size is large, showing that our algorithm is capable of generating the seed set of good quality efficiently. Meanwhile, our algorithm yields competitive results without knowledge of diffusion models, especially when diffusion model is PIC. Also the DRL method tends to be much more time consuming than the others.

\noindent
\textbf{Conclusion.}
Analyzing both the influence spread and the computational time, we conclude that our framework efficiently delivers a competitive seed set only slightly inferior to the baselines without the exact knowledge about the diffusion model, and that this performance can be achieved with relatively small diffusion datasets, acceptable training time and fast inference time in the most cases.

\subsection{Relation Learning Verification}
\label{subsec:relation exp}
\begin{figure}[!t]
    \centering
    \subfigure[Facebook - PIC2]{
        \centering
        \includegraphics[width=0.41\linewidth]{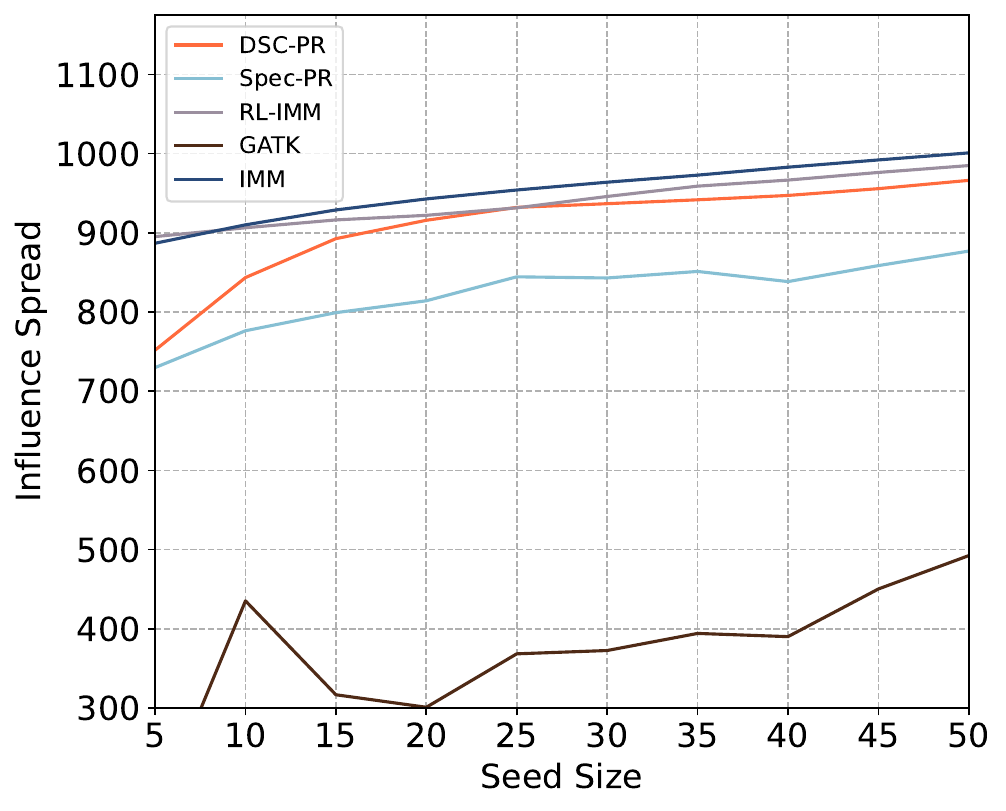}
        \label{fig:comparison_1}
    }
    \subfigure[Twitter - PIC5]{
        \centering
        \includegraphics[width=0.41\linewidth]{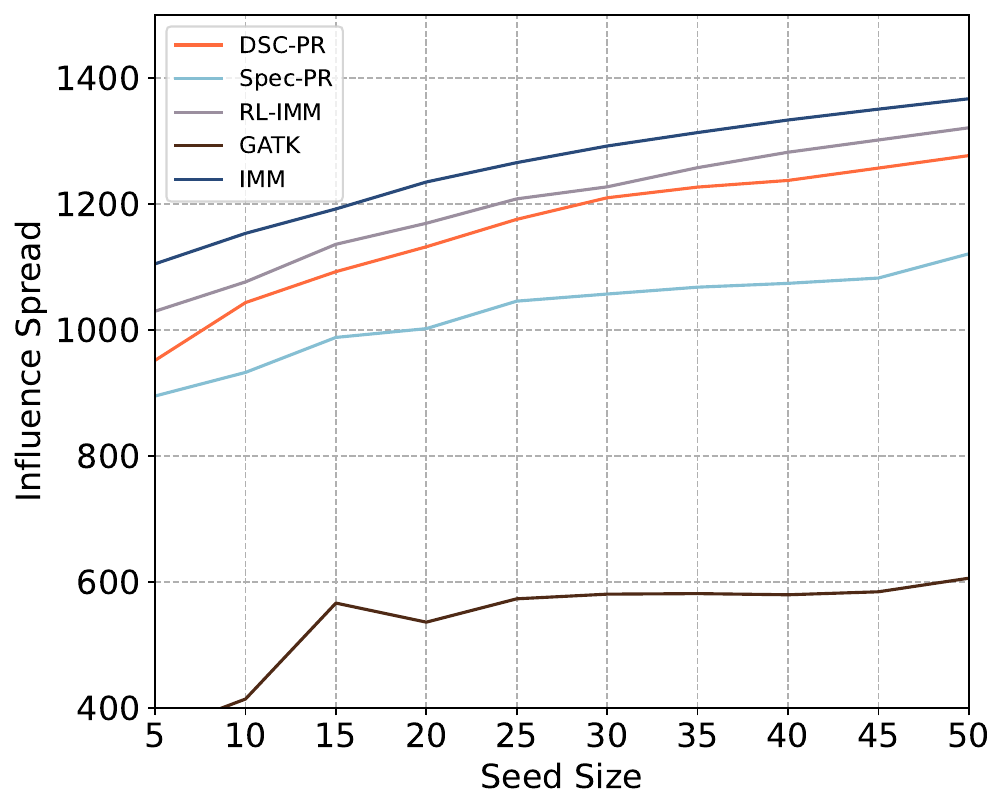}
        \label{fig:comparison_2}
    }
    \centering
    \vspace{-0.3cm}
    \caption{Relation learning verification experiments.}
    \label{fig:comparison}
    \vspace{-0.5cm}
\end{figure}

One of the most critical designs of our DSCom framework is the attention extraction. In order to justify our novel design, we construct the following three comparison experiments separately to show the necessity of the attention extraction, the improvement brought by the relation learning module, and the quality of the learned edge weights. The estimation results of influence spread across different networks are shown in Fig.~\ref{fig:comparison}.

\noindent
\textbf{Clustering Alternative.}
The objective of the attention extraction is to improve the performance of the following clustering algorithm. However, after node embedding, a more general approach is to perform the clustering task within the latent representation space with some Unsupervised Machine Learning algorithms, typically k-means++~\cite{kmeans++}. Then naturally, the seed node within each community can be selected with the one closest to the centroid. We hereby name this approach as GAT-k-means, for short, GATK. Empirical results in Fig.~\ref{fig:comparison} show that the GATK can seriously degrade the algorithm's performance, which verifies the necessity of the attention extraction design.

\noindent
\textbf{Ablation Study.}
To demonstrate the improvement brought by the relation learning module, we perform an ablation study by removing the relation learning module from our framework. For comparison, we choose the same centrality measure as the one of DSCom, and we name this comparison algorithm as Spectral-PageRank, for short, Spec-PR. The performance of Spec-PR is inferior to our D-PR, indicating the improvement of our learning module is significant.

\noindent
\textbf{Weight Evaluation.}
In addition, it would also be interesting to verify that our learned attention coefficients can effectively represent the closeness of the neighbor pairs. To this end, we run the state-of-the-art approximation algorithm, IMM, with the learned parameters. The intuition behind it is that if the algorithm has well approximated the correct diffusion parameters, the state-of-the-art algorithm should deliver a solution almost as good as the IMM with known diffusion parameters. This test algorithm is named Relation-Learning IMM, for short RL-IMM. The experimental results suggest the approximation algorithm that uses the learned weights, RL-IMM, does not degrade too much with respect to the one given the correct model, meaning that the relationship between the neighbor pairs has been well estimated with our attention extract method. We also point out that though RL-IMM is superior to our framework, it is not applicable to our general data-driven problem formulation defined in Def.~\ref{def:data IM}, since the IMM algorithm is only designed for specific diffusion models.

Given these observations, we may safely conclude that the relation learning module is helpful and the attention extraction mechanism is well designed in learning the relationship between nodes.

\section{Conclusion}
\label{sec:conclusion}

Regarding the drawbacks brought by the current statistical approach of data-driven IM problem, in this paper, we reformulate this problem such that it no longer restricted to any specific diffusion pattern and leverage node attributes to estimate the strength of the connections. Targeting this problem, we propose a ML-based framework DSCom to perform the seed selection in a heuristic way. It mainly incorporates two critical designs, the attention coefficient extraction and the community detection with spectral clustering. Experiments on real-world datasets prove that our framework remains competitive to the state-of-the-art approximation algorithm with the disadvantage of not knowing the exact diffusion model. DSCom can deliver a seed set of good quality within an acceptable training time and relatively limited diffusion chains. Empirical results also prove the necessity and the efficiency of our attention extraction design in the relation learning module. Thus, our DSCom does learn the closeness relationship between nodes from the diffusion chains, and we can combine it with some heuristic strategies to successfully achieve the goal of maximizing influence, which does not depend on any pre-defined diffusion model.

%
%
%
%
%
%
%
%
%
%
%
%
\bibliographystyle{splncs04}
\bibliography{DSCom}

\end{document}